# Atomically thin boron nitride: a tunnelling barrier for graphene devices

Liam Britnell[1], Roman V. Gorbachev[2], Rashid Jalil[2], Branson D. Belle[2], Fred Schedin[2], Mikhail I. Katsnelson[3], Laurence Eaves[4], Sergey V. Morozov[5], Alexander S. Mayorov[1], Nuno M. R. Peres[6,7], Antonio H. Castro Neto[7], Jon Leist[8], Andre K. Geim[1,2], Leonid A. Ponomarenko[1], Kostya S. Novoselov[1*]

[1]*School of Physics & Astronomy, University of Manchester, Manchester M13 9PL, UK*
[2]*Manchester Centre for Mesoscience & Nanotechnology, University of Manchester, Manchester M13 9PL, UK*
[3]*Institute for Molecules and Materials, Radboud University of Nijmegen, 6525 AJ Nijmegen, The Netherlands*
[4]*School of Physics & Astronomy, University of Nottingham, Nottingham NG7 2RD, UK*
[5]*Institute for Microelectronics Technology, 142432 Chernogolovka, Russia*
[6]*Departamento de Física, Universidade do Minho, P-4710-057, Braga, Portugal*
[7]*Graphene Research Centre and Department of Physics, National University of Singapore, 2 Science Drive 3, 117542, Singapore*
[8]*Momentive Performance Materials, 22557 West Lunn Road, Strongsville, OH 4407*

*We investigate the electronic properties of heterostructures based on ultrathin hexagonal boron nitride (h-BN) crystalline layers sandwiched between two layers of graphene as well as other conducting materials (graphite, gold). The tunnel conductance depends exponentially on the number of h-BN atomic layers, down to a monolayer thickness. Exponential behaviour of I-V characteristics for graphene/BN/graphene and graphite/BN/graphite devices is determined mainly by the changes in the density of states with bias voltage in the electrodes. Conductive atomic force microscopy scans across h-BN terraces of different thickness reveal a high level of uniformity in the tunnel current. Our results demonstrate that atomically thin h-BN acts as a defect-free dielectric with a high breakdown field; it offers great potential for applications in tunnel devices and in field-effect transistors with a high carrier density in the conducting channel.*

Hexagonal boron nitride (h-BN) has great potential for use as the dielectric layer in functional heterostructure devices which exploit the remarkable properties of graphene (1-4). The combination of graphene and h-BN opens up the exciting possibility of creating a new class of atomically-thin multilayered heterostructures (5,6). Graphene and h-BN share the same crystal structure and have very similar lattice constants but, unlike graphene, h-BN is an insulator with a large energy bandgap of 6 eV (7,8). Most previous studies have focused on the use of thick layers of BN as a substrate for graphene electronics (7,9,10) or as a dielectric in experiments on coupled 2D electron gases (11). HBN has also been used as a barrier for tunneling experiments: thick hBN (>6 layers) was sandwiched between two graphene layers (12) and thinner layers of this material (down to monolayer) were used in gold/hBN/gold tunneling devices (13). Extension of these results to graphene and graphite tunneling diodes with a monolayer of hBN as a tunnelling barrier is of considerable fundamental interest and has the potential for new applications such as devices for flexible electronics, especially as the layer thickness can, in principle, be controlled with atomic layer precision.

Here we investigate the electronic properties of tunnel diodes in which h-BN acts as a barrier layer between a variety of different conducting materials, such as graphene, graphite and gold. We demonstrate that a single atomic layer of h-BN acts as an effective tunnel barrier and that the transmission probability of the h-BN barrier decreases exponentially with the number of atomic layers. The current-voltage characteristics of these devices show a linear *I-V* dependence at low bias and an exponential dependence at higher applied voltages. We use conductive atomic force microscopy (C-AFM) to measure the tunnel current through h-BN terraces of different thickness and find that the tunnel current on a particular terrace is spatially uniform and defect-free.

To investigate the electronic properties of the BN barriers, we fabricated several types of device, in the form of the following sandwich structures: Au/BN/Au; graphene/BN/graphene; and graphite/BN/graphite. For the Au/BN/Au samples we fabricated gold stripes of typical width 2 μm from a 5 nm Ti + 50 nm Au metallic bilayer deposited on a Si/SiO$_2$ substrate in which the oxide layer was 100 nm thick. Flakes of BN were then deposited on the metallic stripes using a micromechanical cleavage technique (1). The flake thickness was characterised by a combination of optical contrast (8), Raman spectroscopy and AFM methods. BN crystallites of different thicknesses which overlapped the gold contacts were identified (8) and top contacts (5nm Ti/50nm Au) were deposited by electron beam or laser-writing lithography and electron gun evaporation.

We used an alternative technique to fabricate the graphene/BN/graphene and graphite/BN/graphite devices. To form the bottom electrode, micromechanical cleavage (1,5) was used to deposit narrow flakes of graphene (or graphite) on a $SiO_2$ substrate. Where necessary, the flakes were narrowed by reactive plasma etching through a PMMA mask. Similarly prepared and characterised BN crystals were deposited on the graphene (or graphite) flakes using a dry-transfer technique (7,10). The top graphene (or graphite) electrode was transferred by the same method; (it could be shaped prior to transfer to achieve the desired overlapping area). Figure 1 shows representative images of our devices.

We measured the *I-V* characteristics of our samples over a range of temperatures. The most reliable results were obtained with graphene or graphite as contact layers: for these devices, the measured current scaled accurately with the device area. However, gold contacts are somewhat less reproducible. We attribute this difference to the atomic flatness of the graphene and graphite layers. In contrast, the BN tunnel barrier may delaminate mechanically from the rough surface of the metallic layer, thus leading to a change in the active surface area of the device and to a reduction of the tunnel current. As shown in Figs. 2 and 3, the *I-V* curves of all our devices are linear around zero bias but have an exponential dependence on V at higher biases. The zero-bias conductivity for each type of device scales exponentially with the BN barrier thickness and is of the order of 1 $k\Omega^{-1}$ $\mu m^{-2}$ for a monolayer BN sandwiched between two graphite electrodes, decreasing to approximately 0.1 $G\Omega^{-1}$ $\mu m^{-2}$ for devices with 4 BN atomic layers, see Fig. 4. The Au/BN/Au and graphene/BN/graphene devices demonstrate similar behaviour but with a larger spread of estimated current densities due to surface roughness and the effects of residual doping respectively.

Figure 3 demonstrates the exponential dependence of the tunnel current on bias voltages above 0.5 V. Interestingly, for the graphene/BN/graphene and graphite/BN/graphite devices the exponential dependence does not scale with barrier thickness; it is approximately independent of barrier thickness, which indicates that the tunnel current is dominated by the so-called tunneling density of states. This behaviour can be understood in terms of a simple theoretical model (14,15) in which the tunnel current is given by (15)

$$I(V) \propto \int dE\, DoS_B(E) DoS_T(E-eV) T(E)[f(E-eV) - f(E)]$$

where $f(E)$ is the Fermi distribution function, $DoS_{B(T)}(E)$ is the density of states in the top (bottom) electrode, $T(E)$ is the transmission probability at the given energy. At low temperatures the difference of the Fermi functions restricts the relevant energy $E$ integral to $\mu < E < \mu + eV$ where $\mu$ is the chemical potential. The above formula assumes that there is no in-plane momentum conservation, which is most likely to be the case of realistic graphene-hBN interfaces. There are several possible mechanisms for elastic scattering at the interface and, in particular, unavoidable fluctuations of the mass term due to the lattice mismatch. In our modeling we utilized a realistic density of states for graphene (graphite) and assumed the hole tunneling with effective mass $m=0.5m_0$ ($m_0$ is the free electron mass) and the height of the tunneling barrier $\Delta \approx 1.5$ eV (16,17).

In order to examine the uniformity and defect density in the BN insulating barrier layer, we carried out a separate set of experiments using an atomic force microscope (AFM) tip as the top electrode and graphite as the lower electrode, see Fig. 3. When the AFM tip was maintained at the same position (with negligible thermal drift), the *I-V* characteristics were very similar to those obtained for the graphite/BN/graphite samples. By comparing the tunnel currents obtained on the graphite/BN/graphite and the graphite/BN/tip devices, we were able to estimate the active area of the tip to be $\sim 10^3$ nm$^2$, in reasonable agreement with the number expected for metal-coated AFM tips with nominal radius of 10 nm. For the AFM tip experiments, we observed little variation in the amplitude of the current for repeated ramps in bias voltage. This suggests that the effective area of the contact varied little over the course of the experiment. However, one would anticipate that the tip area could change due to the mechanical forces exerted by the sample surface. Several similar tips were used on a range of samples: the resulting *I-V* curves followed the same trend for all tip and sample combinations, with only small discrepancies in current due to differing tip shapes/areas.

We also scanned graphite/BN sandwich devices with conducting AFM tips in contact mode with a constant bias applied between the graphite layer and tip, Fig 5. The resistance map is very uniform, with only small variations of about 10%. For nearly all of the devices we observed no evidence of pinholes or defects, even though areas of tens of square microns were scanned. This demonstrates that BN can be used as an atomically thin, high-quality insulator.

In conclusion, we have demonstrated a method for fabricating devices with atomically-thin tunnel barriers by cleaving a few layer flakes from an h-BN crystal and transferring them without contamination to a suitable contact electrode layer, namely graphene, graphite or Au mounted on a Si/SiO$_2$ substrate. A variety of devices can be fabricated by using different combinations of these materials for the upper and lower electrodes. Our measurements of the electron tunnel current through the barrier demonstrate that the BN films act as a good tunnel barrier down to a single atomic plane. Current-voltage measurements were made for different BN thicknesses, from 1 to 4 atomic layers. The *I-V* curves are linear at low bias but take on an exponential dependence at higher bias (> 0.4 V). The tunnel current is exponentially dependent on the BN barrier thickness, as expected for quantum tunneling. Our C-AFM measurements indicate that the tunnel current is uniform when the tip is scanned across micron scale areas of a flat BN atomic layer, indicating an absence of pinholes and defects in the BN crystal. We conclude that h-BN is a high quality, ultrathin, low dielectric constant barrier material, which has great potential for use in novel electron tunneling devices and for investigating strongly-coupled and narrowly-separated electrodes of different compositions (graphene, graphite or metal). Of particular interest is the possibility of studying the electronic properties of two closely-spaced graphene layers separated by a BN barrier of one or more atomic layer thicknesses.

**Acknowledgements**

This research was supported by the European Research Council, European Commision FP7, Engineering and Physical Research Council (UK), the Royal Society, U.S. Office of Naval Research, U.S. Air Force Office of Scientific Research, and the Körber Foundation.

**Figures**

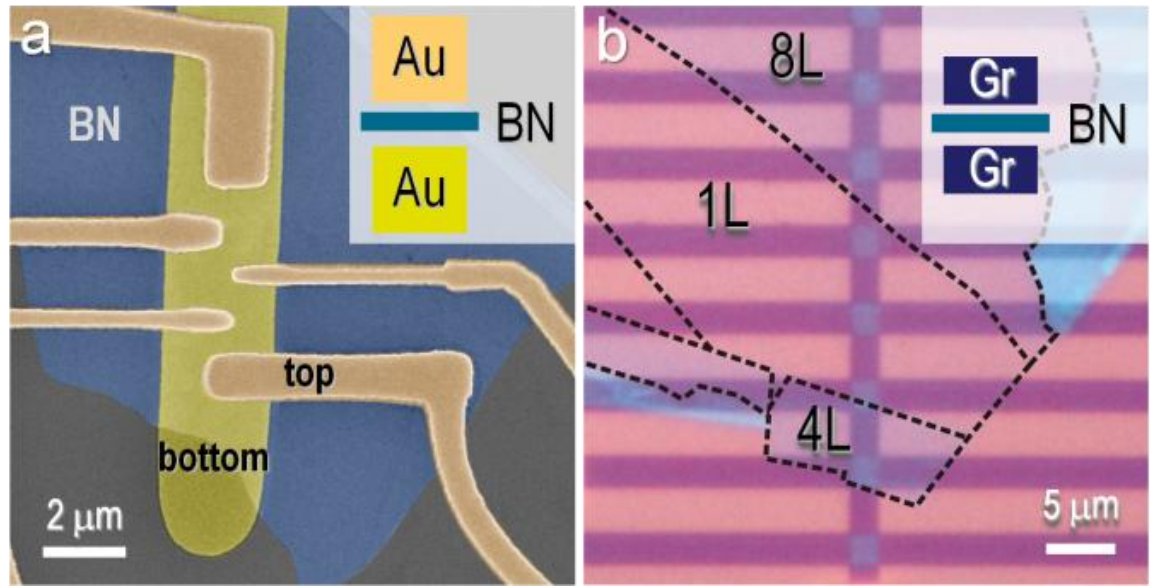

Fig. 1 Micrographs of two of our devices. A. An SEM micrograph of one of our Au/BN/Au devices (false colours). The bottom golden electrode is partly covered with 6 layers of BN (bluish area of irregular shape). Five top golden electrodes of different areas are then deposited on top of BN. Inset: schematic representation of the device. B. An optical image of one of our graphite/BN/graphite devices. The bottom graphite layer is shaped by reactive plasma etching into several stripes of 2.5μm width (horizontal purple lines). Thin BN layers have high transparency so the edges of BN crystallites with different numbers of layers are marked by black lines. Crossing of the top graphite layer (vertical purple line, 2.5μm in width) forms several tunneling junctions with different thicknesses of BN. Inset: schematic representation of the device.

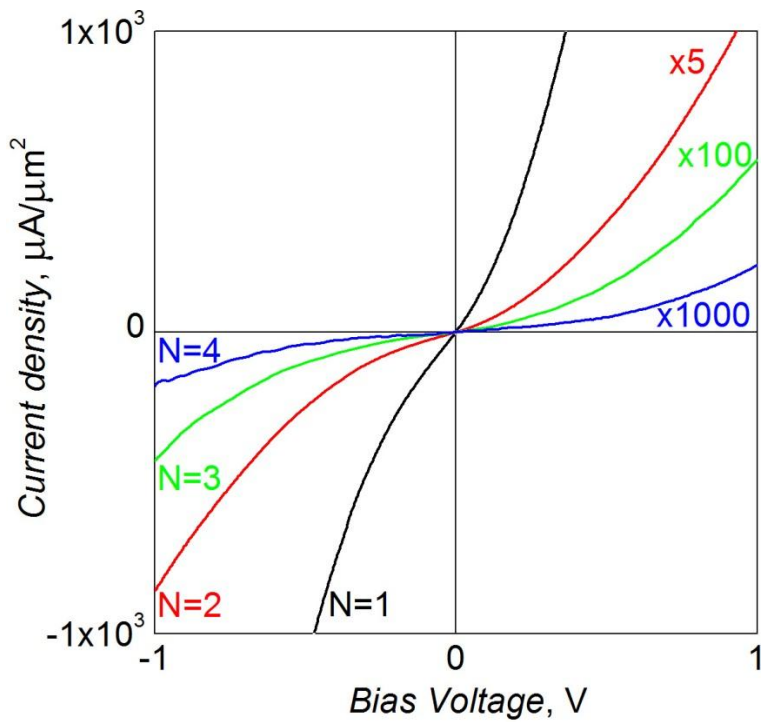

Fig. 2 Characteristic *I-V* curves for graphite/BN/graphite devices with different thicknesses of BN insulating layer. Black curve – monolayer of BN, red – bilayer, green – triple layer, blue – quadruple layer. Note the different scale for the four curves. Current was normalised by the realistic area of the tunneling barrier, which ranged 2-10 μm$^2$ depending on the particular device.

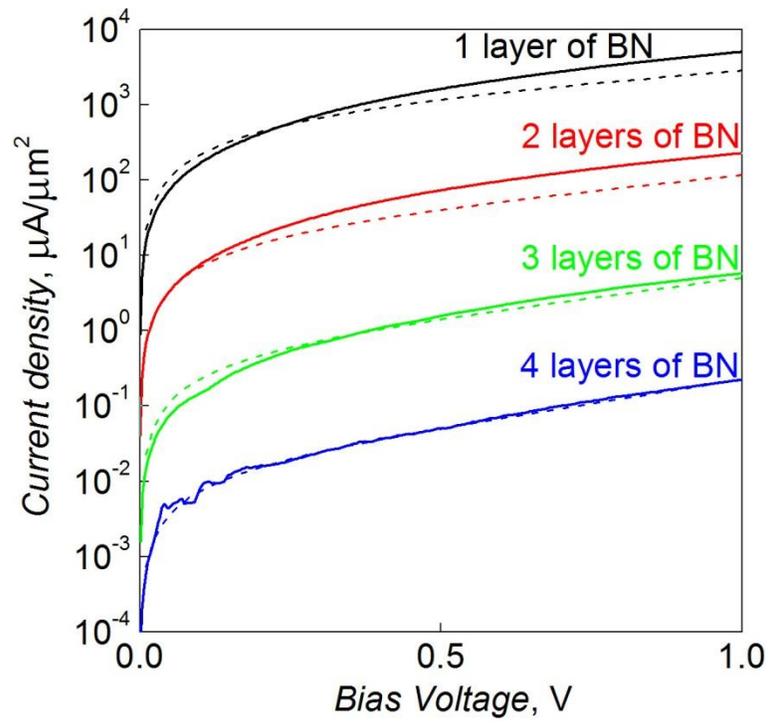

Fig. 3 Same as Fig. 2, just in log scale. Solid curves – experimental data, dashed lines – our modelling. Only one fitting parameter for all four curves was used.

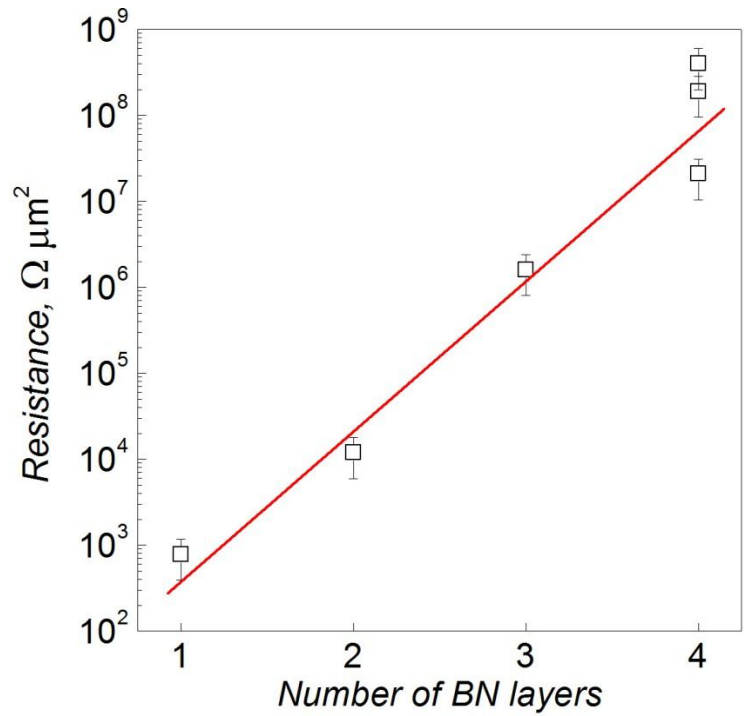

Fig. 4 Exponential dependence of zero-bias resistance on the thickness of BN separating graphite electrodes (1 to 4 layers of BN, 0.3-1.3 nm). Resistance is normalised to the area, which ranged 2-10μm$^2$ depending on the particular device.

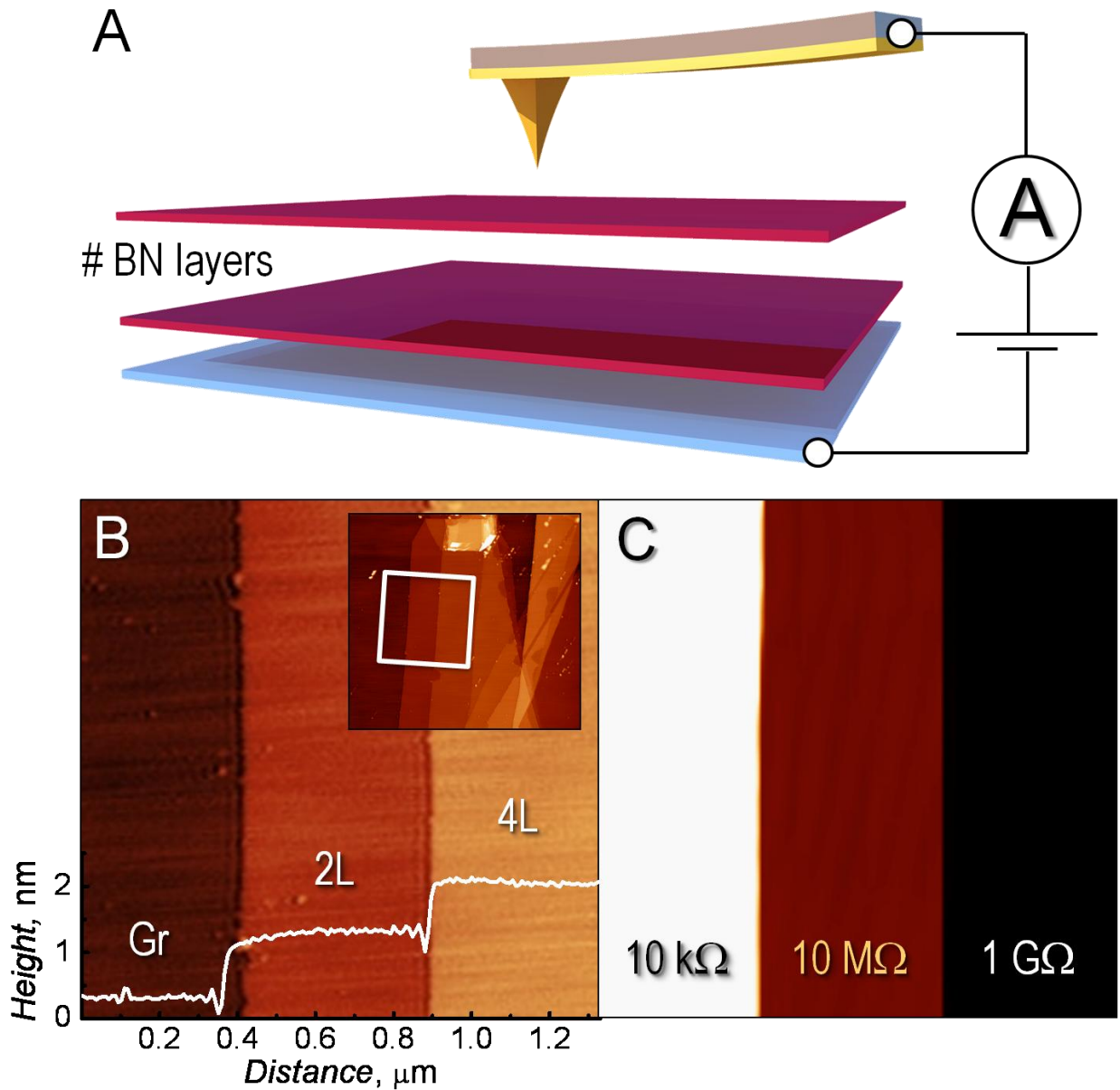

Fig. 5 Local mapping of the tunneling current through BN terraces over micron scale regions by C-AFM measurements. A. Schematic representation of the set-up. Graphite (the blue plane) is covered by several layers of BN. A conductive AFM probe is used to scan the surface and the tunneling current recorded as a function of the tip position. B. Topography measured by tapping-mode AFM, the inset shows the overall region and the height profile shows that BN has a low

roughness (<1 nm RMS). C. The leakage current measured at the same position as topography in B. There is little variation of leakage current on regions of the same thickness suggesting a lack of pin holes or other defects that would lead to spikes in the measured current.